\documentclass[11pt,a4paper]{article}

\usepackage[english]{babel}
\usepackage[utf8]{inputenc}

\usepackage{epsf}
\usepackage{cite}
\usepackage{graphicx}
\usepackage{subcaption} 

\usepackage{amsmath}
\usepackage{amssymb}
\usepackage{mathrsfs}
\usepackage[normalem]{ulem}
\usepackage{mathbbol}
\usepackage{multirow}
\usepackage{slashed}
\usepackage{amsfonts}
\usepackage{bbding}
\usepackage{array}
\usepackage[
colorlinks=true
,urlcolor=black
,anchorcolor=black
,citecolor=black
,filecolor=black
,linkcolor=black
,menucolor=black
,linktocpage=true
,pdfproducer=medialab
,pdfa=true
]{hyperref}

\usepackage{color} 
\usepackage[left=2cm,right=2cm,top=2cm,bottom=2cm]{geometry}
\usepackage{tikz}
\usetikzlibrary{shapes,arrows,positioning,automata,backgrounds,calc,er,patterns}
\usepackage[compat=1.1.0]{tikz-feynman}
\usepackage{contour}
\allowdisplaybreaks
\begin{document}
\begin{center}
\vspace*{1mm}
\vspace{1.3cm}
{\Large\bf
Taming flavour violation in the Inverse Seesaw
}

\vspace*{1.2cm}

{\bf Jonathan~Kriewald $^\text{a}$ and Ana~M.~Teixeira $^\text{b}$ }

\vspace*{.5cm}
$^\text{a}$ {Jožef Stefan Institut, Jamova Cesta 39, P. O. Box 3000, 1001 Ljubljana, Slovenia}

\vspace*{.2cm}
$^\text{b}$ Laboratoire de Physique de Clermont Auvergne (UMR 6533), CNRS/IN2P3,\\
Univ. Clermont Auvergne, 4 Av. Blaise Pascal, 63178 Aubi\`ere Cedex,
France

\end{center}

\vspace*{5mm}
\begin{abstract}
\noindent
The Inverse Seesaw mechanism remains one of the most attractive explanations for the lightness of neutrino masses, allowing for natural low-scale realisations. 
We consider the prospects of a simple extension via 3 generations of sterile fermions - the so called ISS(3,3) - in what concerns numerous lepton flavour observables. 
In order to facilitate a connection between the Lagrangian parameters and low-energy data, we systematically develop new parametrisations of the Yukawa couplings.
Relying on these new parametrisations to explore the parameter space, we discuss the complementary role of charged lepton flavour violation searches in dedicated facilities, as well as in lepton colliders (FCC-ee and $\mu$TRISTAN). 
Our results reveal the strong synergy of the different indirect searches in probing the distinct flavour sectors of the model. In particular, we show that in the absence of radiative decays $\ell_\alpha\to\ell_\beta\gamma$, sizeable rates for $Z$-penguin dominated observables could hint at a non-trivially mixed and non-degenerate heavy spectrum.

\end{abstract}

\section{Introduction}
Oscillation data (i.e. the smallness of neutrino masses and the pattern of leptonic mixings) remains one of the most pressing open issues in particle physics, signalling a clear departure from the Standard Model (SM). 
Among the numerous New Physics (NP) models which have been put forward to address the problem of neutrino mass generation, certain constructions offer the appealing possibility of being realised at low-energies, opening the door to direct searches for the new resonances, and/or to indirect signals. The latter emerge as a consequence of new contributions to both SM-like observables (including for example electroweak precision tests) and to SM-forbidden processes (as is the case of charged lepton flavour violation transitions and decays).  

The Inverse Seesaw mechanism (ISS)~\cite{Schechter:1980gr, Gronau:1984ct, Mohapatra:1986bd} consists in a variant of the type I Seesaw~\cite{Minkowski:1977sc,Yanagida:1979as,Glashow:1979nm,Gell-Mann:1979vob,Mohapatra:1979ia}, in which two species of sterile fermions are added to the SM particle content,  $X$ and $\nu_R$. The relevant terms in the Lagrangian are given by
\begin{equation}\label{eq:ISS:lagrangian}
    \mathcal L_\text{ISS}\, =\, -Y^D_{ij} \,\overline{L_i^c}\,\widetilde H \,\nu_{Rj}^c - M_R^{ij}\, \overline{\nu_{Ri}}\, X_j - \frac{1}{2}\mu_R^{ij}\, \overline{\nu_{Ri}^c}\,\nu_{Rj} - \frac{1}{2} \mu_X^{ij}\, \overline{X_i^c}\, X_j + \text{H.c.}\,.
\end{equation}
From this Lagrangian, the mass matrix of the neutral fermions in the basis $(\nu_L, \nu_R^c, X)$ can be cast as
\begin{equation}
    M_\text{ISS} 
    = \begin{pmatrix}
    \mathbb{0} & m_D & \mathbb{0}\\
    m_D^T & \mu_R & M_R\\
    \mathbb{0} & M_R^T & \mu_X
    \end{pmatrix} 
    \,,\label{eqn:mass_matrix}
\end{equation}
with $m_D = v\,Y_D/\sqrt{2}$, in which $v$ denotes the electroweak (EW) vacuum expectation value.
By setting $\mu_{X,R}\to 0$, one recovers total lepton number conservation as a global symmetry of the Lagrangian in Eq.~\eqref{eq:ISS:lagrangian}, and having small $\mu_{X,R}$ becomes technically natural in the sense of 't Hooft~\cite{tHooft:1980xss,Hettmansperger:2011bt}.
In the limit of small $\mu_{X,R}\ll m_D \ll M_R$, one can thus perturbatively (block-)diagonalise the mass matrix in Eq.~\eqref{eqn:mass_matrix} and obtain to leading order
\begin{eqnarray}
    m_\nu \simeq m_D\, (M_R^{-1})^T\, \mu_X\, M_R^{-1}\, m_D^T \equiv U_\text{PMNS}^\ast\, m_\nu^\text{diag}\, U_\text{PMNS}^\dagger\,,
    \label{eq:ISS:lightmasses}
\end{eqnarray}
for the light neutrino masses.
The mass spectrum of the heavy sterile states is instead strongly restricted, since these combine to form approximately degenerate pseudo-Dirac pairs, as a consequence of the small lepton number breaking via $\mu_X$ (with the mass splittings proportional to $\mu_X$).
Notice that the Majorana mass term $\mu_R$ is absent from Eq.~\eqref{eq:ISS:lightmasses}, as it only appears in higher orders in the seesaw expansion (and in loop corrections to the neutrino mass matrix~\cite{Dev:2012sg}); thus, in the interest of simplicity it will be henceforth neglected.
In what follows we will consider a ``symmetric'' realisation of the ISS in which $n_R = n_X = 3$ generations of heavy sterile fermions are added to the SM particle content, the so-called ISS(3,3)\footnote{For a detailed discussion of the most minimal ISS realisations, see~\cite{Abada:2014vea}.}, leading to square matrices $\mu_X, M_R, m_D$.

Due to the ``double-suppression'' of simultaneously having a large $M_R$ and a small $\mu_X$, the mass scale of the physical heavy states can be lowered to the TeV-scale while retaining $\mathcal O(1)$ Yukawa couplings.
Thus, the ISS can have a very rich phenomenology with potentially sizeable contributions to a plethora of low-energy (flavour) observables~\cite{Abada:2014vea,Abada:2014nwa,Abada:2014kba,Abada:2014cca,Arganda:2015naa,Abada:2015oba,Abada:2018qok,Abada:2024hpb}, (precision) observables at the $Z$-pole~\cite{Arganda:2014dta,DeRomeri:2016gum,2207.10109,2307.02558} as well as potentially detectable collider signatures~\cite{Das:2012ze,Arganda:2015ija,Das:2016hof,Cai:2017mow,Pascoli:2018heg,Abada:2022wvh}.
Consequently, it is desirable to have a clear connection between the Lagrangian parameters and low-energy data, and ideally express them in terms of masses and mixings which can be related to physical observables.

A first attempt can be made relying on a modified Casas-Ibarra parametrisation~\cite{Casas:2001sr}, which allows to directly incorporate neutrino oscillation data into the Yukawa couplings by means of the Pontecorvo-Maki-Nakagawa-Sakata (PMNS) matrix and the measured atmospheric and solar mass squared differences.
However, as we proceed to point out, the connection to other low-energy data such as various lepton universality tests, proves to be rather difficult due to the non-linear nature of the involved matrix equations.
Here, we systematically develop alternative parametrisations that directly encode low-energy data from universality tests and oscillation data into the Yukawa couplings and (Majorana) mass matrices.
We further demonstrate their usefulness in disentangling distinct sources of flavour violation, that is flavour-violation from active-sterile mixing, and from mixings exclusively amongst the heavy states. 
This is done through a dedicated phenomenological analysis of various low-energy charged lepton flavour violation (cLFV) observables, cLFV and lepton flavour universality violation (LFUV) $Z$-pole observables, as well as the potential impact of cLFV searches at a $\mu^+ e^-$ ($\mu$TRISTAN) collider.

As we proceed to show, the parametrisations we propose here include ``hidden'' parameters (in this case the mixing amongst the heavy states); just like for the Casas-Ibarra parametrisation, these parameters can have a significant impact on the predictions of low-energy observables.
However, and in contrast to the Casas-Ibarra parametrisation, those we propose here exhibit no hyperbolic dependence on these parameters, and so  their impact for the observables is  less severe.
Furthermore, we study in detail the impact of the ``hidden'' parameters in the case of non-degenerate heavy spectra, ultimately pointing out that only observables that depend on $Z$-penguin transitions can be affected, and can in turn serve as probes of the mixing in the heavy spectrum.

Our manuscript is organised as follows.
In Section~\ref{sec:ISSpara}, we systematically develop new parametrisations of the ISS(3,3) from simple algebraic arguments. 
In Section~\ref{sec:cLFVEW} we describe the observables and their experimental status, which will be used in the phenomenological analysis of Section~\ref{sec:pheno}. We finally conclude in Section~\ref{sec:concs}.

\section{Parametrising the ISS(3,3): incorporating low-energy data}
\label{sec:ISSpara}

As mentioned before, a low-energy mechanism of neutrino mass generation as the ISS is expected to lead to abundant contributions to flavoured observables and to EW precision probes. 
In order to explore the flavour content of such classes 
of models, phenomenological studies frequently rely on parametrisations of the flavour structures of the model
(which typically ensure that neutrino oscillation data is accommodated).

As a possible first approach to study the phenomenology of the ISS, one can begin by considering a modified Casas-Ibarra parametrisation~\cite{Casas:2001sr}, 
\begin{equation}
  v\, Y_D^T \,=\,   m_D^T = V^\dagger\,\sqrt{M^\text{diag}}\,R\,\sqrt{m_\nu^\text{diag}} \, U_\text{PMNS}^\dagger\,.
    \label{eqn:ISSCI}
\end{equation}
In the above, the Yukawa couplings encode neutrino data, with 
a complex orthogonal matrix, $R$, parametrising the  
additional degrees of freedom. Moreover, in Eq.~(\ref{eqn:ISSCI}) the unitary matrix 
$V$ diagonalises $M = V^\dagger\, M^\text{diag}\,V^\ast$ with $M=M_R \,\mu_X^{-1} \,M_R^T$. 

While the Casas-Ibarra parametrisation allows for a simple (and for the most part numerically stable) access to the ISS(3,3) parameter space that is consistent with neutrino oscillation data, it has significant drawbacks.
Firstly, the ``$R$-matrix'' has no direct physical interpretation, thus leading to ambiguities which cannot be easily resolved.
Secondly, an arbitrary complex {\it orthogonal} matrix - as $R$ - is parametrised via 3 Euler rotations, which naturally leads
to a hyperbolic dependence on the imaginary parts of the mixing angles. In turn, this can translate into a poor behaviour of  
the numerical 
sampling of the parameter space, as one quickly runs into non-perturbative regimes (see e.g.~\cite{Casas:2010wm} for attempts to cure this behaviour).
Finally, 
due to the ``entanglement'' of the PMNS and $R$-matrix flavour structures in $m_D$ (cf. Eq.~(\ref{eqn:ISSCI})), it becomes very hard (if not impossible) 
to disentangle different directions in ``flavour-violation-space''. It further renders complying with broad classes 
of low-energy data (cLFV, EW, LFUV, ...) a complicated and not very transparent exercise. 

As originally proposed in~\cite{FernandezMartinez:2007ms}, 
the deviations from unitarity of the $3\times 3$ would-be PMNS block of the unitary leptonic mixing matrix offer a convenient means of encoding the constraints from low-energy data. 
The so-called ``$\eta$-matrix'' is defined as
\begin{equation}
\label{eq:defPMNSeta}
U_\text{PMNS} \, \to \, \tilde U_\text{PMNS} \, = \,(\mathbb{1} - \eta)\, 
U_\text{PMNS}\,.
\end{equation}
Perturbatively diagonalising the ISS(3,3) mass matrix, one can derive an approximate expression for $\eta$ as
\begin{equation}
    \eta \simeq \frac{1}{2} \,m_D^\ast \,(M_R^{-1})^\dagger \,(M_R^{-1}) \,m_D^T\,.
    \label{eqn:eta}
\end{equation}
It is important to notice that after integrating out the new heavy degrees of freedom, the deviation from unitarity induces a potentially lepton flavour (universality) violating $d=6$ operator at tree-level~\cite{Broncano:2002rw,2107.12133}
\begin{equation}
    \mathcal L_{d=6} = \frac{i}{2}\eta_{\alpha\beta}\left[(\bar L_\alpha \gamma_\mu  L_\beta)(\phi^\dagger \overset{\leftrightarrow}{D}_\mu \phi) - (\bar L_\alpha \gamma_\mu \tau^i L_\beta)\,(\phi^\dagger \overset{\leftrightarrow}{D_\mu}{}^{\!\!i}\phi)\right]\,.
    \label{eqn:d6}
\end{equation}

\bigskip
In view of this brief discussion, it becomes desirable to have a parametrisation of the neutrino mass matrix that allows for a calculable connection between the Lagrangian parameters and low-energy data beyond neutrino oscillations. In what follows, we will explore several possible avenues to do so, describing the underlying algebraic approach, as well as the most relevant phenomenological consequences.

Let us first notice that 
inserting the Casas-Ibarra parametrisation for $m_D$ into the definition of $\eta$, and subsequently inverting $R$ might be technically possible but is highly impractical.
Another possibility would be to reverse-engineer a parametrisation for $M_R$ that would allow controlling the ``non-unitarity'' of $\tilde U_\text{PMNS}$; nonetheless, this would imply losing control of the heavy mass scale ($M_R$), which would also be impractical from a phenomenologist's point of view (especially if one desires to infer information on the scale of NP from low-energy data).
In the case of invertible $M_R$ and $m_D$, an alternative parametrisation has been put forward in~\cite{1405.4300} (see also~\cite{Garnica:2023ccx} and~\cite{2307.02558} for modified versions), in which oscillation data is encoded in the lepton number violating term $\mu_X$ rather than in $m_D$.
Considering here the ISS(3,3), for any invertible $M_R$ and $m_D$ one can then write
\begin{equation}\label{eq:parametrisation:muX1}
 \mu_X\, = \,M_R^T \,m_D^{(-1)} \,U_\text{PMNS}^\ast\, \mathrm{diag}(m_{\nu_1}, m_{\nu_2}, m_{\nu_3}) \,U_\text{PMNS}^\dagger \, (m_D^T)^{(-1)} \,M_R\,.
\end{equation}
In principle, this allows retaining full control over the heavy mass scale and the flavour structure of $m_D$, while the Majorana mass $\mu_X$ becomes intrinsically related 
to the scale of the active neutrino masses.
This is, from a phenomenologist's point of view, a very useful approach to {\it systematically} explore low-energy implications of the ISS.
Although the only condition on $M_R$ is that it be invertible, 
for the purpose of this work
we will focus our attention on possible flavour structures of $m_D$.

In~\cite{1405.4300} and~\cite{1607.05257} the authors propose to reverse-engineer textures for $m_D$ which are expected to lead to maximal effects (e.g. cLFV)
along certain ``flavoured-directions'': for instance, aiming at maximising $\mu\tau$-flavour violating observables while evading the strong constraints stemming from 
$e-\mu$ sector flavour observables.
For this purpose, it was proposed to 
construct fixed textures for $m_D$, only varying an overall factor; likewise, the heavy masses are taken to be degenerate, and the mass scale $M_R$ is only varied through an overall factor.
The parametrisation of $m_D$ (where $m_D$ is for simplicity assumed to be real) used in~\cite{1405.4300,1607.05257} is thus given by
\begin{equation}
    m_D \, = \, v\,  A \cdot \mathcal O\,,
    \label{eqn:maximal}
\end{equation}
where $A$ is an invertible lower triangular matrix and $\mathcal O$ a real orthogonal matrix. Notice that should 
$M_R$ be diagonal and universal, then $\mathcal O$ trivially disappears from the expression of $\eta$ (cf. Eq.~(\ref{eqn:eta})), 
and does not contribute to flavour violation in this minimal scenario.
The texture of $A$ is then chosen to optimally explore certain flavoured configurations (such as the example mentioned above - i.e., maximising 
flavour violation in the $e\tau$ and $\mu\tau$ directions, while strongly suppressing $e\mu$ transitions).

The authors of~\cite{Garnica:2023ccx} further notice that it can be useful to re-scale $m_D$ with $M_R/v$, so that one parametrises the active-sterile mixing rather than the Yukawa couplings. 
This was further exploited in~\cite{2307.02558}, where 
the goal was to find regions of the parameter space which allowed maximising lepton flavour universality violation while retaining control over lepton flavour violation.
The desired features can be automatically ensured by parametrising $m_D$ as follows
\begin{equation}
    m_D \,= \,\mathrm{diag}(y_1, y_2, y_3) \cdot\mathcal V \cdot M_R^T\,,
\end{equation}
in which $\mathcal V$ is a unitary matrix. Upon insertion of 
this parametrisation in Eq.\eqref{eqn:eta} one is readily led to 
\begin{equation}
    \eta \,= \,\mathrm{diag}(y_1^2, y_2^2, y_3^2)\,,
\end{equation}
so that flavour violation can only appear via a non-trivial $\mathcal V$, and only in the case of non-degenerate eigenvalues of $M_R$. 
Furthermore, constraints from low-energy data can be trivially incorporated by appropriately fixing $y_i$.

\bigskip
In the present study we aim at generalising the work that has been done in~\cite{1405.4300,1607.05257,Garnica:2023ccx,2307.02558}, with the goal of disentangling PMNS (i.e. oscillation data) from additional beyond the SM (BSM) sources of flavour violation, relying on a simple algebraic approach (using the properties of invertible square matrices). 
Throughout our discussion we will always focus on the ISS(3,3) realisation - even if not explicitly mentioned.

We begin by noticing that the only requirement on $m_D$ is that it be invertible, i.e. that its determinant be different from 0.
Any matrix has a polar decomposition, which is unique in the case of an invertible square matrix.
We can thus write $Y_D = m_D/v$ in full generality as
\begin{equation}
    Y_D^\text{polar} \,= \,P \cdot \mathcal U \,,
\end{equation}
in which $P$ is a positive definite hermitian matrix $P^\dagger = P$ and $\mathcal U$ is unitary $\mathcal U^\dagger\,\mathcal U = \mathbb{1}$.
The generally 9 complex (or equivalently 18 real) free parameters of $m_D$ are encoded in
3 real angles and 6 phases in $\mathcal U$, 3 real diagonal elements of $P$, and 3+3 real parameters in the complex off-diagonal elements of $P$.
Inserting this definition of $Y_D$ into Eq.~\eqref{eqn:eta}
(with the re-scaling of $Y_D$ with $M_R$), 
one quickly finds 
\begin{equation}
    \eta \,= \,\frac{1}{2}\,P^\ast \,P^T\,,
\end{equation}
such that $P$ can now be written as 
\begin{equation}
    P \,= \,\sqrt{2} \,\eta^\frac{1}{2}\,,
\end{equation}
where $\eta^\frac{1}{2}$ is a {\it hermitian} matrix square-root\footnote{Via the eigenvalue decomposition one can easily show that all square roots of a positive definite hermitian matrix are also hermitian and positive definite.} of $\eta$.
The matrix square-root can be found numerically through the Schur method or eigenvalue decomposition, or analytically with the help of the Cayley-Hamilton theorem, as recently derived in~\cite{2403.07756}.
Following the eigenvalue decomposition, it is further clear that in order to ensure the invertibility of $m_D$, $\eta^\frac{1}{2}$ and therefore $\eta$ have to be invertible (i.e. non-singular), which restricts the values $\eta$ can take.
Due to its phenomenological origin as a ``deviation from unitarity'' of the PMNS matrix, the off-diagonal elements of $\eta$ are not completely arbitrary and are subject to Schwarz inequalities given by~\cite{Fernandez-Martinez:2016lgt,Blennow:2023mqx}
\begin{equation}
    |\eta_{ij}| \leq \sqrt{\eta_{ii}\,\eta_{jj}}\,.
    \label{eqn:schwarz}
\end{equation}
Notice that the off-diagonal elements of $\eta$ can have complex phases; only their magnitudes are constrained by Eq.~\eqref{eqn:schwarz}.
In order to automatically comply with low-energy data we can parametrise $\eta$ as
\begin{equation}
    \eta = \begin{pmatrix}\eta_{ee} &\sqrt{\eta_{ee}\,\eta_{\mu\mu}}\, a \exp(i \delta_{12}) & \sqrt{\eta_{ee}\,\eta_{\tau\tau}}  \,b \exp(i \delta_{13}) \\
    \sqrt{\eta_{ee}\,\eta_{\mu\mu}} \,a \exp(-i \delta_{12}) & \eta_{\mu\mu} & \sqrt{\eta_{\mu\mu}\,\eta_{\tau\tau}} \,c \exp(i \delta_{23})\\
    \sqrt{\eta_{ee}\,\eta_{\tau\tau}} \,b \exp(-i \delta_{13}) & \sqrt{\eta_{\mu\mu}\,\eta_{\tau\tau}}\, c \exp(-i \delta_{23}) & \eta_{\tau\tau}
    \end{pmatrix}\,,
\end{equation}
in which $0 < \eta_{ee,\mu\mu,\tau\tau} < 1$ and $0\leq a, b,c < 1$.
The non-singularity of $\eta^\frac{1}{2}$ allows deriving  a further condition from the non-vanishing determinant as
\begin{equation}
    \mathrm{det}\eta^\frac{1}{2} \,= \,\sqrt{\mathrm{det}\eta} \,= \,\sqrt{\eta_{ee}\,\eta_{\mu\mu}\,\eta_{\tau\tau}\left(1 - a^2 - b^2 - c^2 + 2 a b c \cos(\delta_{12} - \delta_{13} + \delta_{23})\right)}\, \:\neq \, 0\,.
    \label{eqn:det_condition}
\end{equation}
The (diagonal) entries of $\eta$ can then be fixed to the upper limits derived in global analyses as done in~\cite{Blennow:2023mqx}.

\bigskip
Alternatively, $Y_D$ can be parametrised via the help of the $QR$ decomposition:
recall that any matrix $A$ can be decomposed into a unitary matrix $Q$ and an upper triangular matrix $R$ as $A = Q\, R$ and consequently into a lower triangular matrix as $A^\dagger = L Q^\dagger$ in which $L = R^\dagger$.
Thus  $Y_D$ can be parametrised as
\begin{equation}
    Y_D^\text{QR} = L \,\mathcal U\,,
\end{equation}
in which $\mathcal U$ is unitary and $L$ is lower triangular, with $L_{ii}\neq 0 $ (since $Y_D$ must be invertible).
Inserting this parametrisation into Eq.~\eqref{eqn:eta} (again together with the re-scaling of $Y_D$ with $M_R$) one has
\begin{equation}
    \eta \,= \,\frac{1}{2} L^\ast L^T\,,
\end{equation}
and thus $L$ can be found from the Cholesky decomposition of $\eta$ (which is unique since $\eta$ is positive definite).
It is interesting to notice that this parametrisation corresponds exactly to what was found in~\cite{1405.4300,1607.05257} in the limit of a real $Y_D$, see Eq.~\eqref{eqn:maximal}.

Both $Y_D^\text{polar}$ and $Y_D^\text{QR}$ parametrisations have the advantage that compliance with low-energy data in the form of bounds on $\eta$ can be trivially encoded in $Y_D$.
However, one still needs to compute the Cholesky decomposition or matrix square-root of $\eta$ which can be analytically cumbersome.
We detail in Appendix~\ref{app:cholesky} the analytical Cholesky decomposition of a complex $3\times 3$ matrix and refer to~\cite{2403.07756} for the analytical expression for the 8 matrix square-roots (and their inverses).
Moreover, one can accidentally run into cases in which the condition of Eq.~\eqref{eqn:det_condition} is violated; 
setting the phases $\delta_{ij} = 0$ and $a = b = c = 1$ saturates the Schwarz inequalities of Eq.~\eqref{eqn:schwarz} and leads to a vanishing determinant, which in turn makes $\eta$ singular and thus non-invertible.
Additionally, having a small determinant (e.g. by setting $a,b,c$ to a number slightly smaller than 1) can quickly become problematic for the numerical inversion of $m_D$, and lead to significant loss of precision.

\bigskip
A third option\footnote{One can also parametrise a general invertible square matrix via a $LU$ decomposition, a Schur decomposition or an eigenvalue decomposition with complex eigenvalues. However, the connection of the involved matrices to physical observables is a priori much less clear than in the parametrisations proposed in this study, and thus we will not discuss them here.} relies on a singular value decomposition (SVD) of $Y_D$.
Any complex matrix $A$, in this case a complex square matrix, can be decomposed into 
\begin{eqnarray}
    A \,= \,\mathcal V_1 \,\mathrm{diag}(\sigma_1, \sigma_2,  ..., \sigma_n)\,\mathcal V_2^\dagger\,,
\end{eqnarray} 
in which $\mathcal V_{1,2}$ are unitary matrices and $\sigma_i$ are the singular values of $A$.
Ensuring that $A$ is non-singular (and thus invertible) conditions each of the singular values to be $\sigma_i \neq 0$.
Consequently, the Yukawa couplings can be parametrised as
\begin{equation}
    Y_D^\text{SVD} = \mathcal V_1\,\mathrm{diag}(y_1, y_2, y_3)\,\mathcal V_2^\dagger\,,
    \label{eqn:SVD}
\end{equation}
in which $y_i>0$ without loss of generality\footnote{Contrary to the other two parametrisations we have so far discussed, the parametrisation of Eq.~(\ref{eqn:SVD}), which relies on a singular value decomposition of $Y_D$, can likely be generalised to ``asymmetric'' versions of the Inverse Seesaw making use of the properties of left- and/or right-invertible rectangular $m_D$ and $M_R$ matrices. A dedicated study of this more general case lies outside the scope of the present work.}.
The unitary matrices can each be parametrised via 3 Euler angles, 3 ``Dirac-like'' phases $\delta_{ij}$ and 3 ``Majorana-like'' phases $\varphi_i$ as
\begin{equation}\label{eq:V12par}
    \mathcal V_{1,2} \,= \,O_{23} \,O_{13} \,O_{12}\,\mathrm{diag}(e^{i\varphi_1}, e^{i\varphi_2}, e^{i\varphi_3})\,,
\end{equation}
with $O_{ij}$ given in the usual form; for example $O_{23}$ can be given as
\begin{equation}
    O_{23} = \begin{pmatrix}
        1 &0&0\\
        0& \cos\theta_{23} &\sin\theta_{23}e^{i \delta_{23}} \\
        0 & -\sin\theta_{23}e^{-i \delta_{23}}&\cos\theta_{23}
    \end{pmatrix} \,.
\end{equation}
Notice that out of the 6 Majorana-like phases only 3 linear combinations are physical, as can be seen immediately upon insertion of $\mathcal V_1$ and $\mathcal V_2$ in Eq.~\eqref{eqn:SVD}.
Further rescaling $Y_D$ with $M_R^T/v$, and inserting it into Eq.~\eqref{eqn:eta} allows making the connection to low-energy data via
\begin{equation}
    \eta \,= \frac{1}{2} \mathcal V_1^\ast\, \mathrm{diag}(y_1^2,y_2^2,y_3^2) \mathcal V_1^T\,,
\end{equation}
such that in the case of a trivial $\mathcal V_1 = \mathbb{1}$ one is trivially led to $y_i = \sqrt{2\eta_{ii}}$. 
Furthermore, one can relate $\mathcal V_1$ and $y_i$ to the eigenvalue decomposition of $\eta$, for which the analytical expressions are in general very involved.

Notice that one can also fix $y_i$ for the trivial case, and then vary the angles and phases of $\mathcal V_1$ for the purpose of dedicated numerical scans. 
In this case, the Schwartz inequality is automatically fulfilled, and in fact never saturated as long as $y_i > 0$.
Furthermore, the angles $\theta_{ij}$ in $\mathcal V_1$ have a simple geometrical interpretation of directions in 
``flavour violation space''.
In fact, $\mathcal V_1$ describes the flavour-violating mixing between active generations and the heavy states and leads to off-diagonal entries in $\eta$.
In a similar fashion to what occurred in previous examples for the right-hand side unitary matrices, 
the matrix $\mathcal V_2$ also cancels in $\eta$.

\bigskip
To summarise, the explicit parametrisations of the Yukawa couplings $Y_D$ in terms of $\eta$ are
\begin{eqnarray}
\label{eq:YDparsum:polar}    
Y_D^\text{polar} &=& \frac{\sqrt{2}}{v}(\eta^*)^{\frac{1}{2}} \mathcal V_2 M_R^T \,,\\
\label{eq:YDparsum:QR}    
Y_D^\text{QR} &=& \frac{1}{v} L\mathcal V_2 M_R^T \quad\text{with}\quad \eta  =\frac{1}{2} L^\ast L^T\,,\\
    Y_D^\text{SVD} &=& \frac{1}{v} \mathcal V_1 \mathrm{diag}(y_1, y_2, y_3) \mathcal V_2 M_R^T\quad\text{with}\quad \eta = \frac{1}{2}\mathcal V_1^\ast \mathrm{diag}(y_1^2, y_2^2, y_3^2)\mathcal V_1^T\,.
    \label{eq:YDparsum:SVD}
\end{eqnarray}

In all cases, we stress that due to the presence of $M_R^T$ in the rightmost position in all the above expressions for $Y_D$, the connection to low-energy constraints (encoded in $\eta$) is valid for all possible invertible $M_R$.
Thus, one directly parametrises the mixing between the active and sterile states rather than relying on ``unphysical'' Yukawa couplings.
Moreover, the mixing amongst the heavy states is encoded in $\mathcal V_2$.

In particular, it is worth highlighting that  
taking $\mathcal V_2\neq \mathbb{1}$ to be non-trivial leads to an additional mixing in flavour-violation space, if the usual GIM-like suppression is broken by having non-degenerate heavy states.
This can be understood as follows:
the loop function entering the widths for radiative decays $\ell_\alpha\to\ell_\beta\gamma$ asymptotically tends to a constant for $M_R\gg m_W$~\cite{Ilakovac:1994kj} as
\begin{eqnarray}
   \label{eq:rad-ddecays:full}
   G_\gamma(x) &=& -\frac{x(2x^2 + 5x - 1)}{4(1-x)^3} -
    \frac{3x^3}{2(1-x)^4}\log x\,,\\
     G_\gamma(x)&\xrightarrow[x\gg 1]{}& \frac{1}{2}\,,
     \label{eq:rad-ddecays:limit}
\end{eqnarray}
where $x = M_N^2/m_W^2$ and $M_N$ is the mass of the heavy internal fermion. 
For $Z$-penguins (and for simplicity we take here the part which depends only on one internal fermion mass, and at vanishing momentum transfer), the loop functions asymptotically admit a logarithmic dependence on $x$ as~\cite{Ilakovac:1994kj}
\begin{eqnarray}
F_Z(x) &=& -\frac{5 x}{2(1 - x)} - \frac{5x^2}{2(1-x)^2}\log x\,,\\
    F_Z(x) &\xrightarrow[x\gg 1]{}& \frac{5}{2} - \frac{5}{2}\log(x)\,.\label{eq:Zpenguin:limit}
\end{eqnarray}
The dependence on mixing elements for a non-degenerate heavy spectrum is in general more complicated, and the impact of cLFV bounds on the off-diagonal elements of $\eta$ is model-dependent, as discussed in~\cite{Blennow:2023mqx}.
Consequently, the different parametrisations outlined here will lead to different results for cLFV processes generated by anything other than just dipole operators, due to the different matrix structures that enter $Y_D$ (see Eqs.~(\ref{eq:YDparsum:polar}-\ref{eq:YDparsum:SVD})).
\section{Future cLFV searches and EW precision measurements}
\label{sec:cLFVEW}
As mentioned in previous sections, we will investigate in detail how different parametrisations of the ISS(3,3) allow to better control the contributions for an extensive set of observables, including those sensitive to the violation of lepton flavour universality, and those signalling lepton flavour violation. 
The role of these sets of observables has been widely explored in the framework of SM model extensions via heavy sterile states. For the case of cLFV leptonic transitions (as radiative decays, three-body decays and conversion in nuclei, among others), the corresponding form-factors and loop functions - common to generic variants of type I seesaw - can be found in~\cite{Ilakovac:1994kj,Alonso:2012ji,Abada:2018nio,Abada:2022asx,Riemann:1982rq,Illana:1999ww,Mann:1983dv,Illana:2000ic,Ma:1979px,Gronau:1984ct,Deppisch:2004fa,Deppisch:2005zm,Dinh:2012bp,Abada:2014kba,Abada:2015oba,Abada:2015zea,Abada:2016vzu,Arganda:2014dta}; for the ISS(3,3), cLFV $Z$-boson and Higgs decays have also been investigated~\cite{1405.4300,1607.05257,2207.10109}. Several $Z$-pole observables (including EW precision observables) have been recently evaluated at one-loop level, and a detailed discussion can be found in~\cite{2307.02558}. 
In Table~\ref{tab:cLFV_lep} we summarise the current bounds and future sensitivities for several low- and high-energy cLFV observables. 

A recent study of the prospects of an asymmetric $\mu^+ -e^-$ collider for the discovery of cLFV transitions (induced from the presence of heavy neutral leptons) has shown that a facility as $\mu$TRISTAN can be particularly sensitive to flavour violation in $\tau-\ell$ transitions~\cite{2412.04331}. Here we will also consider the prospects of the ISS(3,3) for such a collider, setting a sensitivity threshold of 10 detected events (with a signal efficiency of $1\%$) for projected luminosities of 100~fb$^{-1}$ and 
1000~fb$^{-1}$.

\renewcommand{\arraystretch}{1.3}
\begin{table}[h!]
    \centering
    \hspace*{-2mm}{\small\begin{tabular}{|c|c|c|}
    \hline
    Observable & Current bound & Future sensitivity  \\
    \hline\hline
    $\text{BR}(\mu\to e \gamma)$    &
    \quad $<3.1\times 10^{-13}$ \quad (MEG II~\cite{MEGII:2023ltw})   &
    \quad $6\times 10^{-14}$ \quad (MEG II~\cite{Baldini:2018nnn}) \\
    $\text{BR}(\tau \to e \gamma)$  &
    \quad $<3.3\times 10^{-8}$ \quad (BaBar~\cite{Aubert:2009ag})    &
    \quad $3\times10^{-9}$ \quad (Belle II~\cite{Kou:2018nap})      \\
    $\text{BR}(\tau \to \mu \gamma)$    &
     \quad $ <4.2\times 10^{-8}$ \quad (Belle~\cite{Belle:2021ysv})  &
    \quad $10^{-9}$ \quad (Belle II~\cite{Kou:2018nap})     \\
    \hline
    $\text{BR}(\mu \to 3 e)$    &
     \quad $<1.0\times 10^{-12}$ \quad (SINDRUM~\cite{Bellgardt:1987du})    &
     \quad $10^{-15(-16)}$ \quad (Mu3e~\cite{Blondel:2013ia})   \\
    $\text{BR}(\tau \to 3 e)$   &
    \quad $<2.7\times 10^{-8}$ \quad (Belle~\cite{Hayasaka:2010np})&
    \quad $5\times10^{-10}$ \quad (Belle II~\cite{Kou:2018nap})     \\
    $\text{BR}(\tau \to 3 \mu )$    &
    \quad $<1.9\times 10^{-8}$ \quad (Belle II~\cite{Belle-II:2024sce})  &
    \quad $5\times10^{-10}$ \quad (Belle II~\cite{Kou:2018nap})     \\
    & & \quad$5\times 10^{-11}$\quad (FCC-ee~\cite{Abada:2019lih})\\
    \hline
    $\text{CR}(\mu- e, \text{N})$ &
     \quad $<7 \times 10^{-13}$ \quad  (Au, SINDRUM~\cite{Bertl:2006up}) &
    \quad $10^{-14}$  \quad (SiC, DeeMe~\cite{Nguyen:2015vkk})    \\
    & &  \quad $2.6\times 10^{-17}$  \quad (Al, COMET~\cite{Krikler:2015msn,COMET:2018auw,Moritsu:2022lem})  \\
    & &  \quad $8 \times 10^{-17}$  \quad (Al, Mu2e~\cite{Bartoszek:2014mya})\\
    \hline
    $\mathrm{BR}(Z\to e^\pm\mu^\mp)$ & \quad$< 4.2\times 10^{-7}$\quad (ATLAS~\cite{Aad:2014bca}) & \quad$\mathcal O (10^{-10})$\quad (FCC-ee~\cite{Abada:2019lih})\\
    $\mathrm{BR}(Z\to e^\pm\tau^\mp)$ & \quad$< 4.1\times 10^{-6}$\quad (ATLAS~\cite{ATLAS:2021bdj}) & \quad$\mathcal O (10^{-10})$\quad (FCC-ee~\cite{Abada:2019lih})\\
    $\mathrm{BR}(Z\to \mu^\pm\tau^\mp)$ & \quad$< 5.3\times 10^{-6}$\quad (ATLAS~\cite{ATLAS:2021bdj}) & \quad $\mathcal O (10^{-10})$\quad (FCC-ee~\cite{Abada:2019lih})\\
    \hline
    \end{tabular}}
    \caption{Current experimental bounds and future sensitivities on relevant cLFV observables. The quoted limits are given at $90\%\:\mathrm{C.L.}$ (Belle II sensitivities correspond to an integrated luminosity of $50\:\mathrm{ab}^{-1}$.)}
    \label{tab:cLFV_lep}
\end{table}
\renewcommand{\arraystretch}{1.}

Likewise, in Table~\ref{tab:obs:EW-LFUV}, we present current experimental measurements and SM predictions for several LFUV and EW observables which will be subsequently discussed in the phenomenological analysis.
 \renewcommand{\arraystretch}{1.3}
\begin{table}[h!]
    \centering
    \hspace*{-2mm}{\small\begin{tabular}{|c|c|c|}
    \hline
    Observable & Exp. Measurement & SM prediction  \\
    \hline\hline
    $R_{\mu e}(Z\to\ell\ell)$ & $1.0001\pm 0.0024$ (PDG~\cite{ParticleDataGroup:2022pth}) & $1.0$~\cite{Freitas:2014hra}\\
    $R_{\tau e}(Z\to\ell\ell)$ & $1.0020\pm0.0032$ (PDG~\cite{ParticleDataGroup:2022pth}) & $0.9977$~\cite{Freitas:2014hra}\\
    $R_{\tau \mu}(Z\to\ell\ell)$ & $1.0010\pm 0.0026$ (PDG~\cite{ParticleDataGroup:2022pth}) & $0.9977$~\cite{Freitas:2014hra}\\
    \hline
    $\Gamma(Z\to e^+e^-)$ & $83.91\pm0.12\:\mathrm{MeV}$ (LEP~\cite{ALEPH:2005ab}) & $83.965\pm0.016\:\mathrm{MeV}$~\cite{Freitas:2014hra}\\
    $\Gamma(Z\to \mu^+\mu^-)$ & $83.99\pm0.18\:\mathrm{MeV}$ (LEP~\cite{ALEPH:2005ab}) & $83.965\pm0.016\:\mathrm{MeV}$~\cite{Freitas:2014hra}\\
    $\Gamma(Z\to \tau^+\tau^-)$ & $84.08\pm0.22\:\mathrm{MeV}$ (LEP~\cite{ALEPH:2005ab}) & $83.775\pm0.016\:\mathrm{MeV}$~\cite{Freitas:2014hra}\\
    \hline
    $\Gamma(Z\to\mathrm{inv.})$ & $499.0 \pm 1.5\:\mathrm{MeV}$ (PDG~\cite{ParticleDataGroup:2022pth})& $501.45\pm 0.05\:\mathrm{MeV}$~\cite{Freitas:2014hra}\\
    \hline
    \end{tabular}}
    \caption{Experimental measurements and SM predictions for several LFUV and EW observables discussed in the phenomenological analysis. All uncertainties are given at 68\% C.L., while for the SM predictions of the universality ratios, the parametric uncertainties are negligible.}
    \label{tab:obs:EW-LFUV}
\end{table}
\renewcommand{\arraystretch}{1.}

Concerning the projections for FCC-ee, in particular for 
$Z$-pole observables~\cite{FCC:2018byv,FCC:2018evy}, it is expected that uncertainties in $R_{\alpha\beta}(Z\to\ell\ell)$ will be reduced to $5\times10^{-5}$, the determination of $\Gamma(Z\to\ell\ell)$ be improved by a factor $\sim 50-500$, while for the $T$-parameter one expects\footnote{Here we take the most conservative estimate in which one assumes no further reduction of intrinsic theoretical uncertainties from missing higher order corrections, as well as no significant improvement of the parametric uncertainties.} $T\lesssim 0.0058$~\cite{deBlas:2019rxi}.

Let us also notice that for other EW observables we will assume a mild improvement by a factor 10 of the associated uncertainties, and display the contours at $95\%$ C.L. upon presentation of our results (solid lines will systematically denote a projection of future uncertainties under the assumption that the central value remains the current one, while dashed lines correspond to assuming that the central experimental value will evolve towards
the SM prediction.

Although we will not offer a detailed discussion here, numerous low-energy processes are sensitive probes to NP sources of lepton flavour universality violation. 
This is the case of leptonic kaon and pion decays, $\tau$-lepton decays, super-allowed beta decays, among many others. 
For simplicity, these constraints can be conveniently encoded in bounds on the diagonal entries of the $\eta$-matrix. 
Below we list the results of a recent analysis~\cite{Blennow:2023mqx} (at $95\%$~C.L.):
\begin{eqnarray}
    \eta_{ee}&\lesssim& 1.4\times 10^{-3}\,,\nonumber\\
    \eta_{\mu\mu} &\lesssim& 1.4\times 10^{-4}\,,\nonumber\\
    \eta_{\tau\tau} &\lesssim& 8.9\times 10^{-4}\,.\label{eqn:etaconstraint}
\end{eqnarray}

\section{Results: exploring the ISS(3,3) parameter space}
\label{sec:pheno}
After the formal discussion of the different techniques to parametrise the new sources of flavour violation that emerge within the Inverse Seesaw, we now proceed to illustrate their comparative potential. Starting from the most simple possibility of a modified Casas-Ibarra parametrisation, we then proceed to consider more sophisticated means of encoding flavour violation, be it between active-sterile states, or in direct connection to the heavy sterile sector. We will further discuss how the ``entanglement'' of sources of flavour violation can be systematically analysed via the parametrisations developed in Section~\ref{sec:ISSpara}.
Throughout this section we assume the light spectrum to be ``normal ordered'', and fix oscillation data to the current central values~\cite{Esteban:2024eli}, further setting the lightest neutrino mass to $10^{-5}\:\mathrm{eV}$.

\subsection{Simplified scenarios}
\begin{figure}
    \centering
    \mbox{\includegraphics[width=0.48\linewidth]{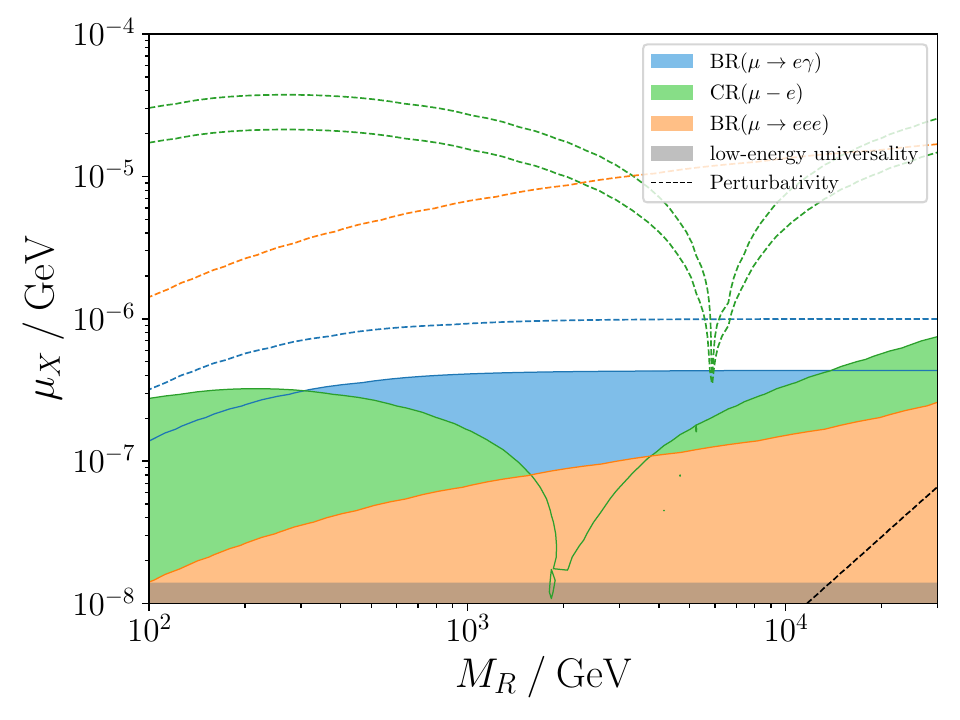}\includegraphics[width=0.48\linewidth]{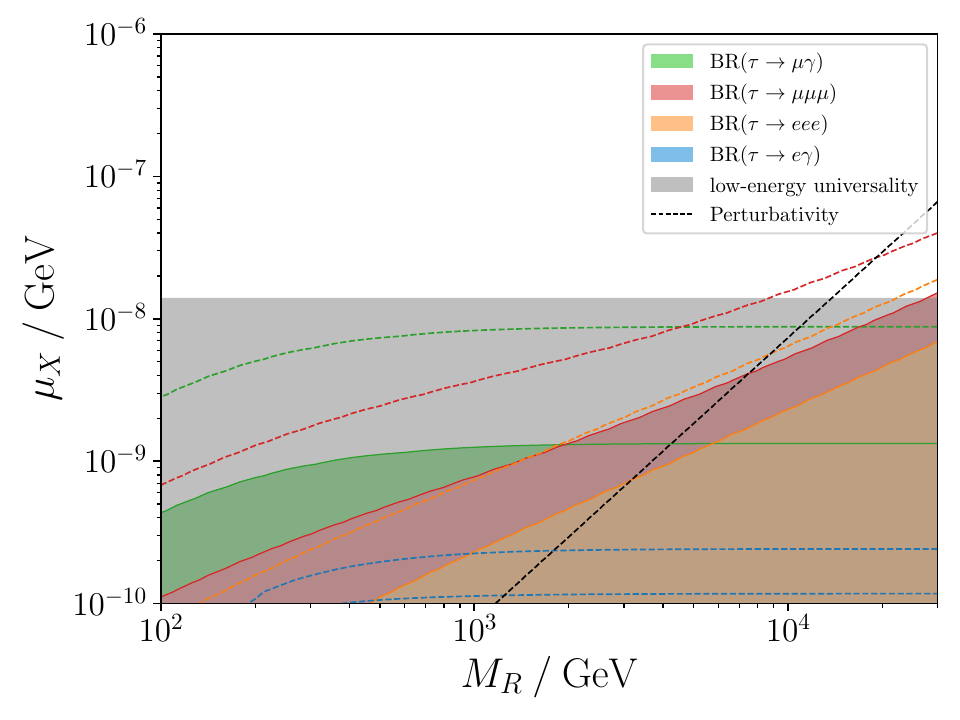}}
    \caption{Casas-Ibarra parametrisation: constraints on the ISS(3,3) parameter space from current bounds and future sensitivities on several cLFV observables, in the $\mu-e$ (left panel) and $\tau-\ell$ (right panel) sectors, together with the bound from fits to low-energy lepton universality observables from~\cite{Blennow:2023mqx}. We have set $R=\mathbb{1}$ and assume $M_R$ and $\mu_X$ to be degenerate.
    Current bounds correspond to shaded regions while future sensitivities are indicated by dashed lines (see Table~\ref{tab:cLFV_lep} for further details). Dark grey regions correspond to exclusion from low-energy universality bounds, while dashed black lines reflect non-perturbative Yukawa couplings.}
    \label{fig:Casas-Ibarra}
\end{figure}

We begin our discussion by a simple - yet illustrative - case, relying in the ISS-modified Casas-Ibarra parametrisation (cf. Eq.~(\ref{eqn:ISSCI})). Taking $M_R$ and $\mu_X$ as diagonal and universal, and setting $R=\mathbb{1}$, we display in Figure~\ref{fig:Casas-Ibarra} several leptonic cLFV processes, separately focusing on $\mu-e$ and $\tau-\ell$ transitions.
Shaded regions denote constraints from current experimental bounds on various cLFV processes, while dashed lines denote the future sensitivity of upcoming (or already running) cLFV experiments.
Note that for neutrinoless $\mu-e$ conversion we display the two most sensitive projections of the COMET and Mu2e experiments (see Table~\ref{tab:cLFV_lep}).
We further display the current upper bound on $\mathrm{Tr}(\eta)$ obtained in~\cite{Blennow:2023mqx} and highlight regimes associated with non-perturbative Yukawa couplings, i.e.  $Y_D^{ij}\geq \sqrt{4\pi}$.

As can be seen in the two plots in Figure~\ref{fig:Casas-Ibarra}, the contributions for $\mu-e$ flavour violating and $\tau$-lepton related observables are strongly entangled with each other and cannot be separately analysed.
 This is a consequence of having a unique ``source'' of flavour violation - the (non-unitary) PMNS matrix.
In order to evade the stringent constraints from data on low-energy universality observables (which do exclude important regions of the $\mu_X - M_R$ parameter space) as well as bounds on $\mu-e$ flavour violating transitions, one could in principle finely tune the complex angles in the $R$-matrix appearing in Eq.~\eqref{eqn:ISSCI}, but there is no clear indication on the path to take.
One could further assume a non-minimal flavour structure in $\mu_X$ and/or $M_R$, but also this approach does not offer a simple connection to $\eta$ due to the non-linear appearance of $\mu_X$ and $M_R$ in Eq.~\eqref{eqn:ISSCI}.

\medskip
Instead of relying on the Casas-Ibarra parametrisation, we take the proposed parametrisation shown in Eq.~\eqref{eqn:SVD}, using a numerical eigenvalue decomposition of $\eta$ to fix $Y_D$ entirely from low-energy universality data (encoded in $\eta_{ii}$) up to $\mathcal V_2$, which we take as trivial, i.e. $\mathcal V_2 = \mathbb{1}$.
In this way, we only parametrise the mixing between active and sterile states while neglecting potential mixings exclusive to  the sterile states.
At this stage, we will assume the heavy spectrum to be degenerate: in other words, we take $M_R$ to be diagonal and universal.

\begin{figure}
    \centering
    \hspace{-0.9cm}\mbox{\includegraphics[width=0.35\linewidth]{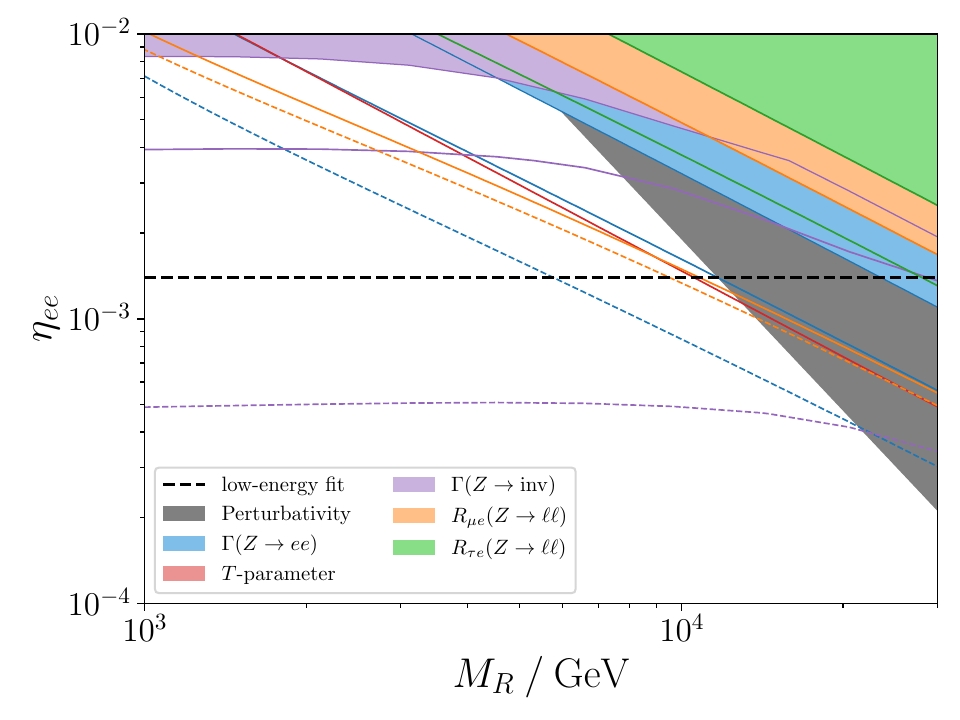}\includegraphics[width=0.35\linewidth]{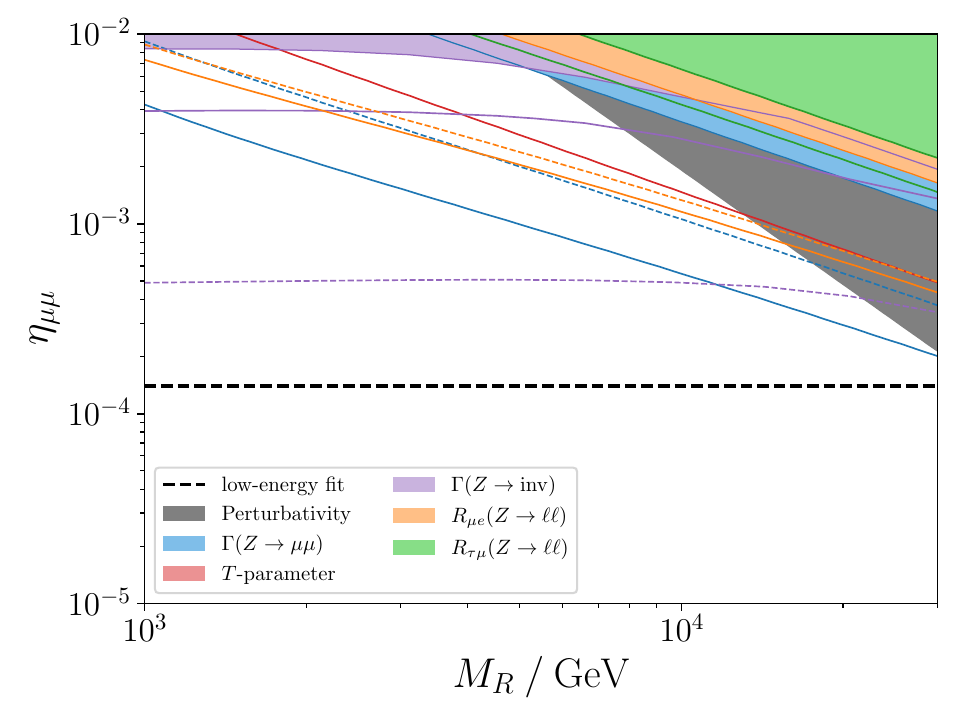}\includegraphics[width=0.35\linewidth]{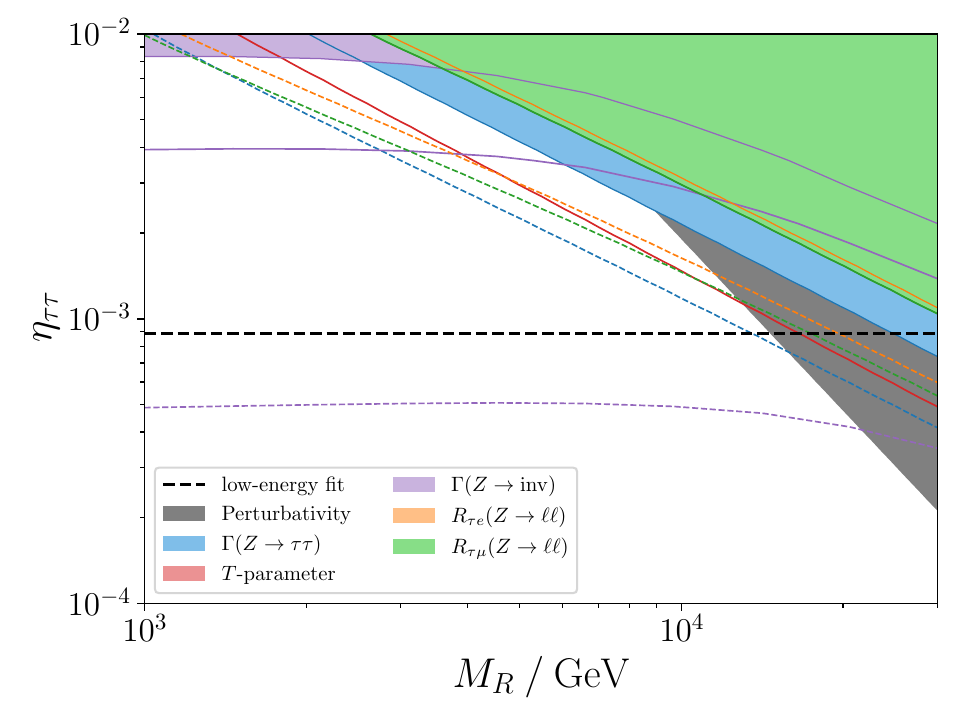}}
    \caption{Constraints from current experimental bounds and future sensitivities from $Z$-pole observables, together with the bound of the low-energy fit on $\eta_{ii}$ of~\cite{Blennow:2023mqx}. From left to right, we consider one $\eta_{ii}$ at a time (the others set to a negligible value) and 
    vary the mass of the heavy states $M_R$ (assumed degenerate). All contours are shown at $95\%$~C.L..}
    \label{fig:eta_diag}
\end{figure}

To begin with, let us then analyse the impact of the mixing of a single heavy pseudo-Dirac pair with the active sector, which can be parametrised via one of the diagonal $\eta_{ii}$.
In this limit (one heavy pair mixing with one active flavour, and degenerate heavy states)
cLFV can only be mediated by light neutrino exchange and is therefore negligible.
In Figure~\ref{fig:eta_diag} we show exclusion contours of current experimental data and future sensitivities of several $Z$-pole observables in the plane spanned by a single $\eta_{ii}$ and the mass of the (degenerate) heavy spectrum $M_R$.
Notice that the contour from the low-energy fit~\cite{Blennow:2023mqx}, cf. Eq.~(\ref{eqn:etaconstraint}), is independent of the heavy mass scale, since only tree-level LFUV observables have been taken into account\footnote{The authors of~\cite{Blennow:2023mqx} in principle also take into account loop-level observables via (tree-level) modifications of the Fermi constant $G_F$, which feed back into electroweak loops that appear, for instance, in corrections to $M_W$. The presence of BSM states in those loops is however neglected, such that the bounds on $\eta$ are independent of the heavy mass scale.}.
The filled contours denote constraints from current data (see Table~\ref{tab:obs:EW-LFUV}), and we also present two distinct projections for future FCC-ee data.
Concerning the latter, and as discussed in the previous section, we assume a rather mild improvement of only one order of magnitude of the uncertainties of $Z$-pole observables due to the unprecedented luminosity of a Tera-$Z$ run at a future $e^+ e^-$ machine.
For the central value of the future measurement, we assume current data (solid lines) or the SM expectation (dashed lines).
In the right-most plot of Figure~\ref{fig:eta_diag}, due to a minor tension between the measurement of $Z\to\tau\tau$ and its SM prediction, the entire plane would be excluded should the central value of $\Gamma(Z\to\tau\tau)$ remain the same. 
(The predictions for all observables shown in Figure~\ref{fig:eta_diag} have been computed relying on the results derived in~\cite{2307.02558}.)
In all plots of Fig.~\ref{fig:eta_diag} it can be seen that the NLO corrections to the $Z$-pole observables induced by the presence of heavy sterile states  might become important in the future, but as far as LEP data is concerned they can safely be neglected.
A more dedicated analysis, also including $W$ and Higgs observables (concerning a potential violation of lepton flavour universality) can be found in~\cite{2307.02558}.

\medskip
We continue by carrying out a complementary approach, focusing on the behaviour of flavour violating observables with respect to the off-diagonal entries of $\eta$.
As outlined in Section~\ref{sec:cLFVEW}, we consider several cLFV observables at low-energies, at the $Z$-pole, as well as at a possible future $\mu$TRISTAN collider, as recently explored in~\cite{2412.04331}.

For heavy and degenerate masses $M_R$, the branching ratios of the cLFV radiative decays are well approximated by
\begin{equation}
    \mathrm{BR}(\ell_\alpha\to\ell_\beta\gamma) \simeq \frac{3\alpha}{2\pi}|\eta_{\alpha\beta}|^2\,,
\end{equation}
due to the aforementioned behaviour of the loop-function entering in the amplitude of the decay (see Eqs.~(\ref{eq:rad-ddecays:full}, \ref{eq:rad-ddecays:limit})).
This is not the case for the $Z$-penguins appearing in $\mu\to 3e$ and neutrinoless $\mu-e$ conversion in muonic atoms, which retain a logarithmic dependence on the heavy mass scale, as illustrated in Eq.~(\ref{eq:Zpenguin:limit}).

In what follows, 
we now fix the diagonal entries of $\eta$ to their maximum values as shown in Eq.~\eqref{eqn:etaconstraint} and vary the degenerate heavy mass scale and one off-diagonal $\eta_{ij}$ at a time (up to its maximum as allowed by the Schwarz inequality, see Eq.~\eqref{eqn:schwarz}), while the others, as well as all phases, have been set to 0. 
\begin{figure}
    \centering
    \hspace{-0.9cm}\mbox{\includegraphics[width=0.35\linewidth]{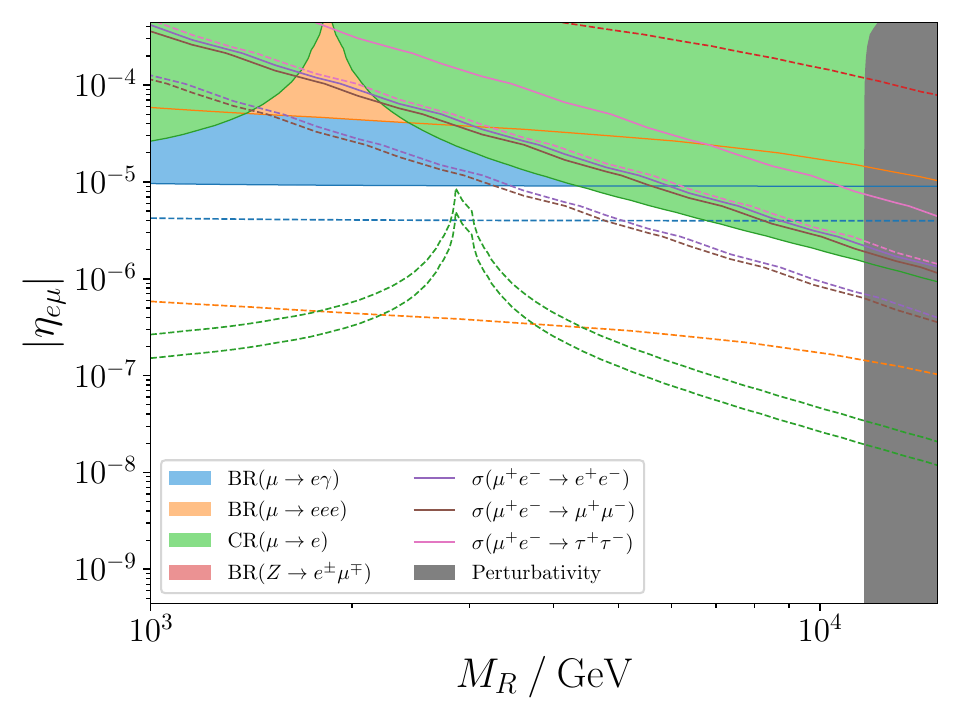}\includegraphics[width=0.35\linewidth]{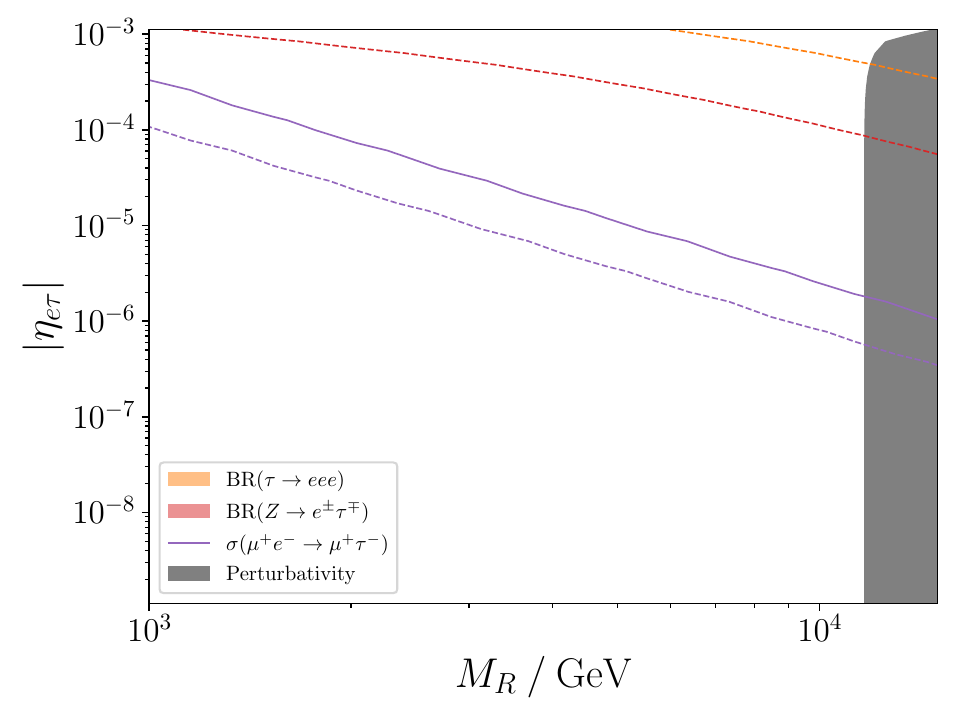}\includegraphics[width=0.35\linewidth]{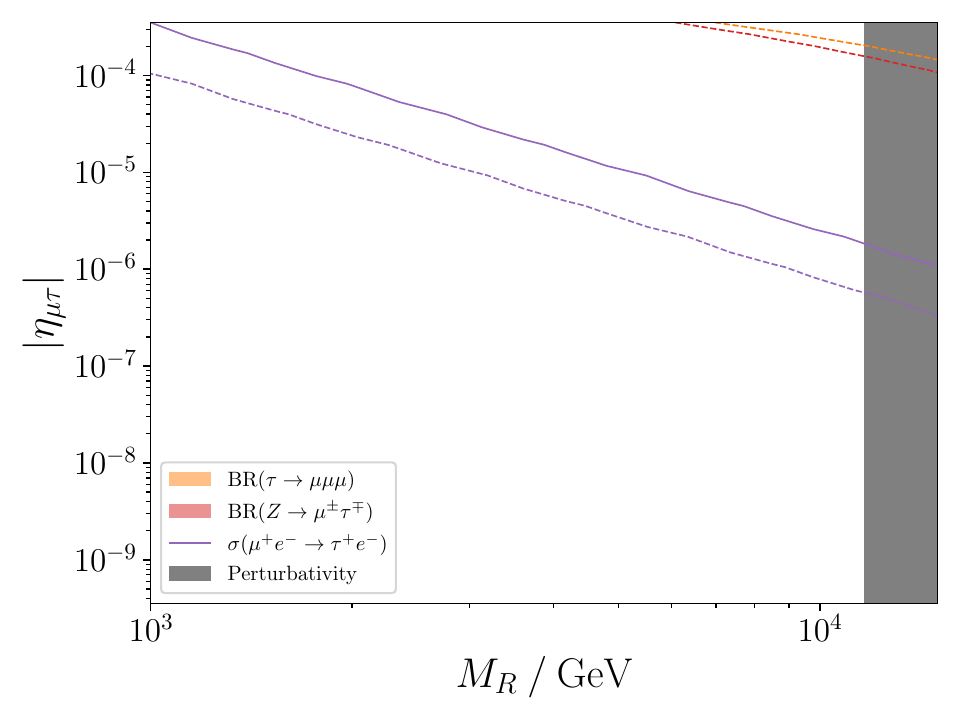}}
    \caption{Constraints from current experimental bounds (filled contours) and future sensitivities (dashed lines) from low-energy and $Z$-pole cLFV observables, together with the projected reach of the $\mu$TRISTAN collider from~\cite{2412.04331}, assuming luminosities of $100\:\mathrm{fb}^{-1}$ (solid lines) and $1\:\mathrm{ab}^{-1}$ (dashed lines). From left to right, we take one off-diagonal $\eta_{ij}$ at a time (with the others set to zero),
    varying the mass of the heavy states $M_R$ (assumed degenerate).
    The diagonal $\eta_{ii}$ have been set to their maximally allowed values as obtained in~\cite{Blennow:2023mqx}.}
    \label{fig:eta_off_diag}
\end{figure}
As shown in the left-most plot of Figure~\ref{fig:eta_off_diag}, the constraints on $\eta_{e\mu}$ derived from either current bounds or future sensitivities (respectively filled contours and dashed lines) on $\mu\to e\gamma$ are nearly independent of 
$M_R$, while constraints from other observables (which are mediated by $Z$-penguins and box topologies) grow stronger for increasing $M_R$, as expected.
We also recover the ``dip'' in $\mu-e$ conversion, which is 
due to a destructive interference between different contributing topologies, and whose position depends on the considered nucleus.
Notice that the current bound for $\mu-e$ conversion was obtained using Gold as a target material, while the two sensitivity projections for the COMET and Mu2e experiments (see Table~\ref{tab:cLFV_lep}) rely on Aluminium, thus shifting the ``dip'' to a different position.

In the remaining plots of Figure~\ref{fig:eta_off_diag} the focus lies on the $\tau-\ell$ sectors.
We display the constraints from future sensitivities of $\tau\to3\ell$, $Z\to \tau^{\pm}\ell^{\mp}$, as well as the projected sensitivities of the preliminary $\mu$TRISTAN analysis of~\cite{2412.04331}.
As already shown in~\cite{2412.04331}, future constraints that can obtained at $\mu$TRISTAN 
on $\eta_{e\tau}$ and $\eta_{\mu\tau}$ from cLFV processes mediated via $\tau-e$ and $\tau-\mu$ $Z$-penguins will strongly outperform bounds stemming both from 
low-energy cLFV processes and $Z$-pole searches at a future Tera-$Z$ factory (even with the most optimistic assumptions about systematic uncertainties); in contrast future $\mu-e$ dedicated facilities will always yield the strongest bounds on $\eta_{e\mu}$.
\subsection{Impact of non-degenerate heavy spectra}
As argued in the previous subsection, the off-diagonal entries $\eta_{ij}$ have a direct connection with radiative decays $\ell_\alpha \to \ell_\beta \gamma$ (which 
are asymptotically independent of the heavy mass scale), in contrast to other cLFV transitions which are dominated by $Z$-penguin exchange.
Moreover, it is important to emphasise that if the heavy spectrum is non-degenerate, the $Z$-penguin mediated processes also depend on a potentially non-trivial flavour structure of the heavy spectrum itself. In turn, this implies that one can no longer ignore the role of $\mathcal V_2$ in the parametrisations derived in Section~\ref{sec:ISSpara}, and it is expected that it will have an impact on the observables.
Throughout the remainder of this section, and to illustrate the impact of a non-degenerate heavy spectrum, we consider a benchmark hierarchy, setting $M_R = \mathrm{diag}(0.9, 1, 1.1) \:M_0$.
Since we aim at isolating the role of a non-trivial $\mathcal V_2\neq \mathbb{1}$, we thus set the off-diagonal $\eta_{ij} = 0$, while keeping the diagonal $\eta_{ii}$ at their maximal values as before (see Eq.~\eqref{eqn:etaconstraint}).
To start with, we vary the overall heavy mass scale $M_0$, and consider the effect of one angle $\sin\theta_{ij}\neq 0$ of $\mathcal V_2$ (with the others set to 0), see Eq.~(\ref{eq:V12par}).

\begin{figure}
    \centering
    \hspace{-0.9cm}\mbox{\includegraphics[width=0.35\linewidth]{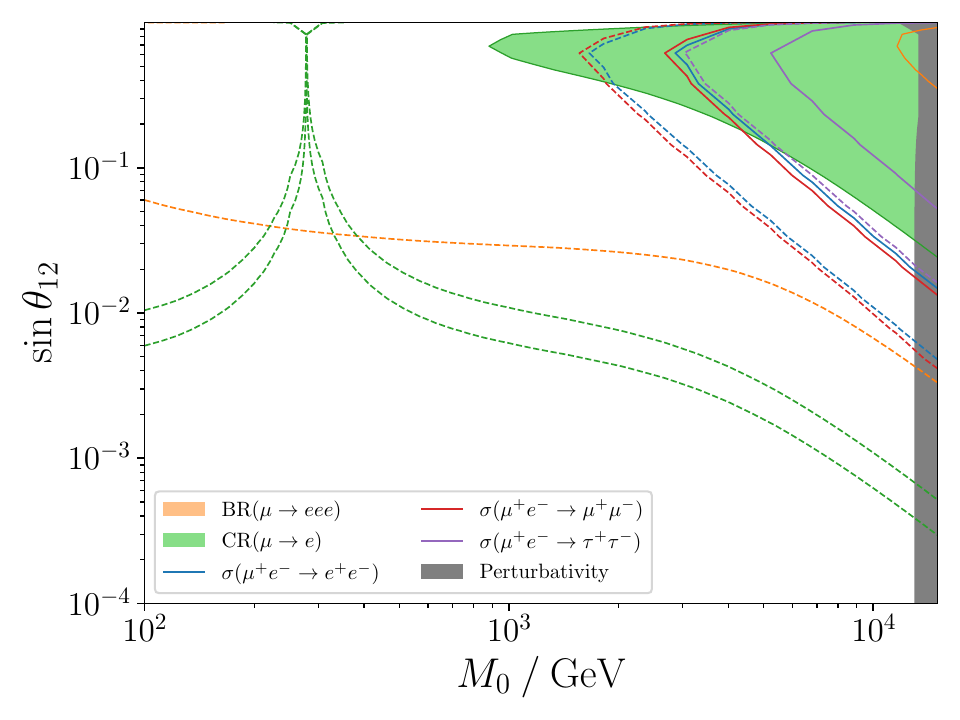}\includegraphics[width=0.35\linewidth]{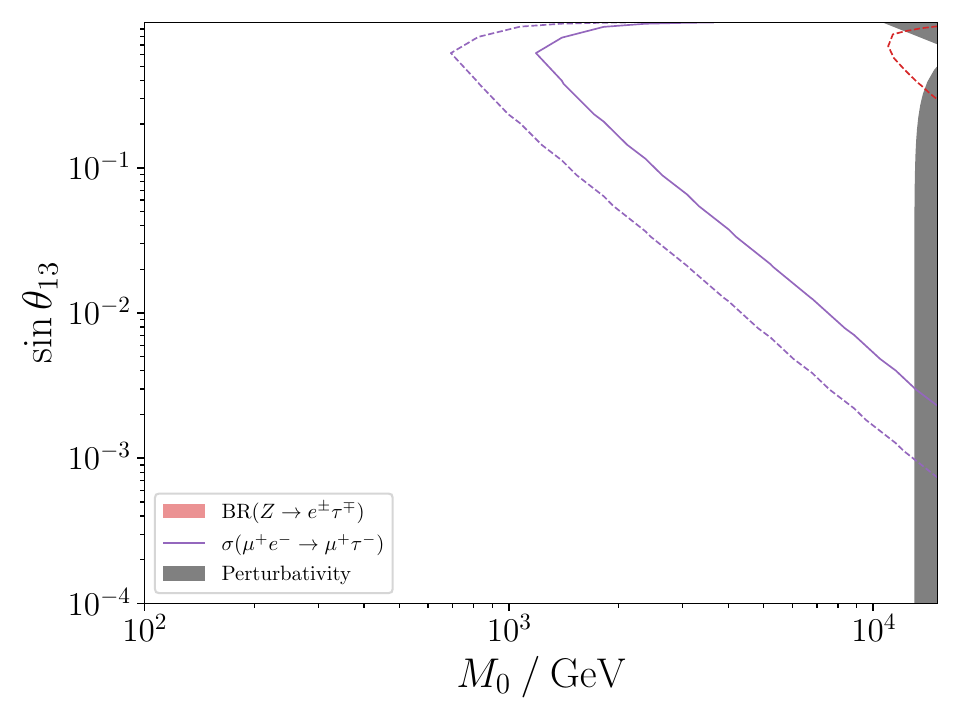}\includegraphics[width=0.35\linewidth]{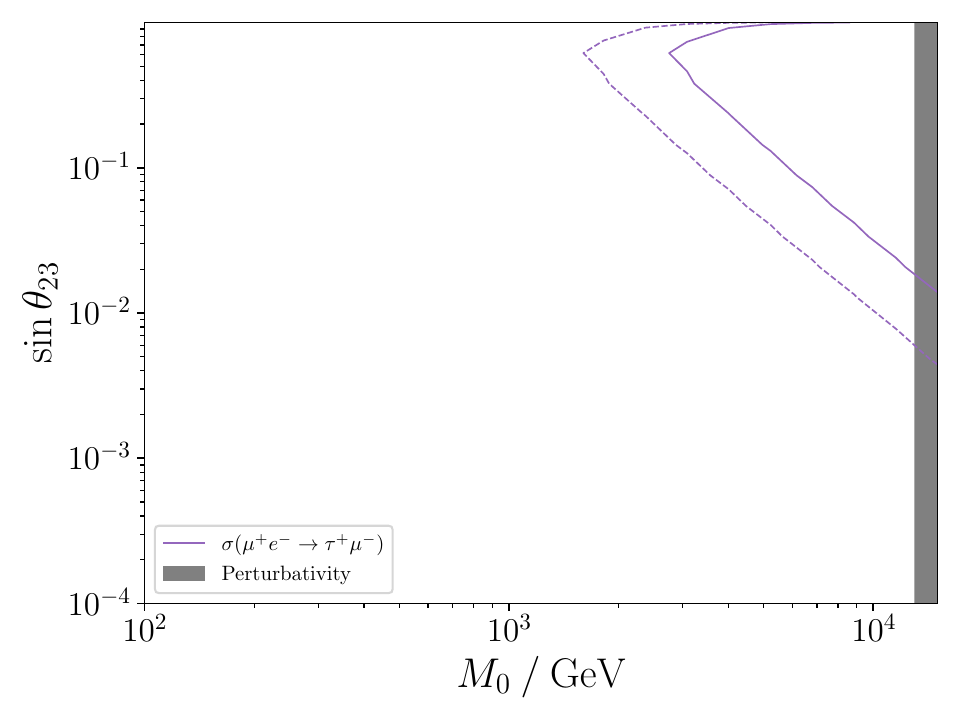}}
    \caption{Constraints from current experimental bounds (filled contours) and future sensitivities (dashed lines) from low-energy and $Z$-pole cLFV observables, together with the projected cLFV reach of a $\mu$TRISTAN collider from~\cite{2412.04331}, assuming luminosities of $100\:\mathrm{fb}^{-1}$ (solid lines) and $1\:\mathrm{ab}^{-1}$ (dashed lines). From left to right, we vary 
    one of the angles of $\mathcal V_2$ (see Eq.~\eqref{eqn:SVD}) at a time with the other angles set to zero, versus
        the mass scale of the heavy states $M_0$ (assuming a non-degenerate spectrum with $M_R = \mathrm{diag}(0.9, 1, 1.1) \:M_0$). The off-diagonal $\eta_{ij}$ have been also set to zero, and the diagonal $\eta_{ii}$ to their maximally allowed values as obtained in~\cite{Blennow:2023mqx}.}
    \label{fig:nd_sij_vs_MR}
\end{figure}

In Figure~\ref{fig:nd_sij_vs_MR} we present the exclusion contours from current data on cLFV (coloured surfaces) as well as the projected reach of dedicated cLFV facilities and that of cLFV searches at $\mu$TRISTAN (dashed lines).
Even if the off-diagonal entries of $\eta$ have been set to 0,
processes such as $\mu\to3e$ and neutrinoless $\mu-e$ conversion in muonic atoms - which receive dominant penguin contributions in the ISS(3,3) - significantly constrain the angle $\sin\theta_{12}$ over a wide range of masses $M_R$.  In this case, $\mu$TRISTAN scattering observables only play a minor role.
(The rate of the radiative decay $\mu\to e\gamma$ is negligible by construction.)
In the remaining plots of Figure~\ref{fig:nd_sij_vs_MR} we recover the leading role of $\mu$TRISTAN for $\tau-\ell$ flavour-violating observables: the strongest (would-be) constraints on $\sin\theta_{13}$ and $\sin\theta_{23}$ would be obtained by $\mu$TRISTAN, while other penguin transitions contributing to low-energy observables and $Z$-pole cLFV are predicted to have very small rates.

\begin{figure}
    \centering
    \mbox{\includegraphics[width=0.48\linewidth]{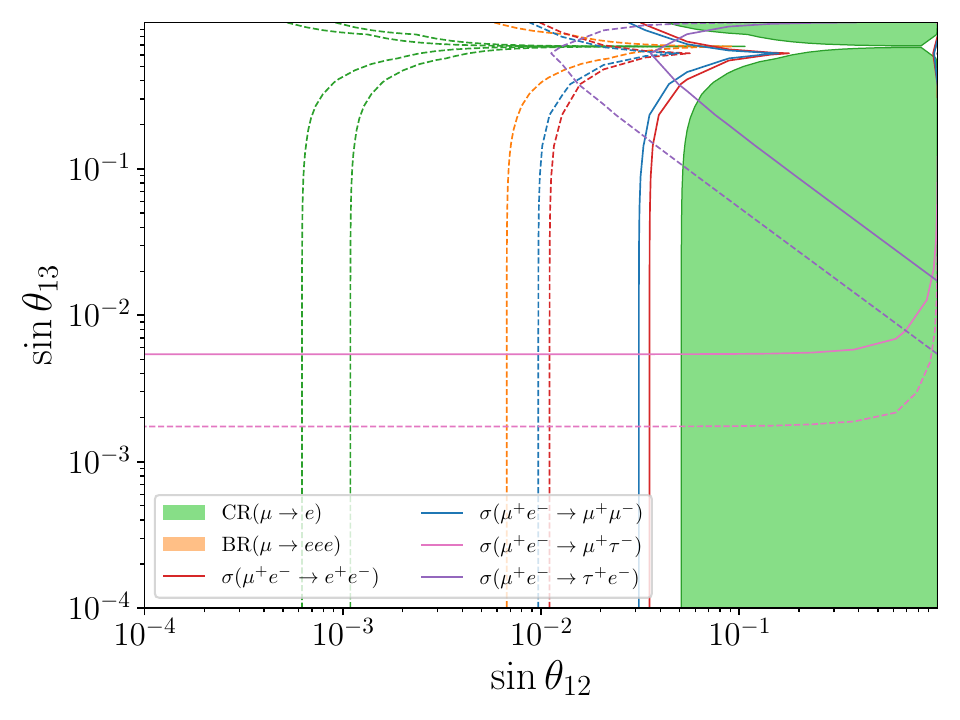}\includegraphics[width=0.48\linewidth]{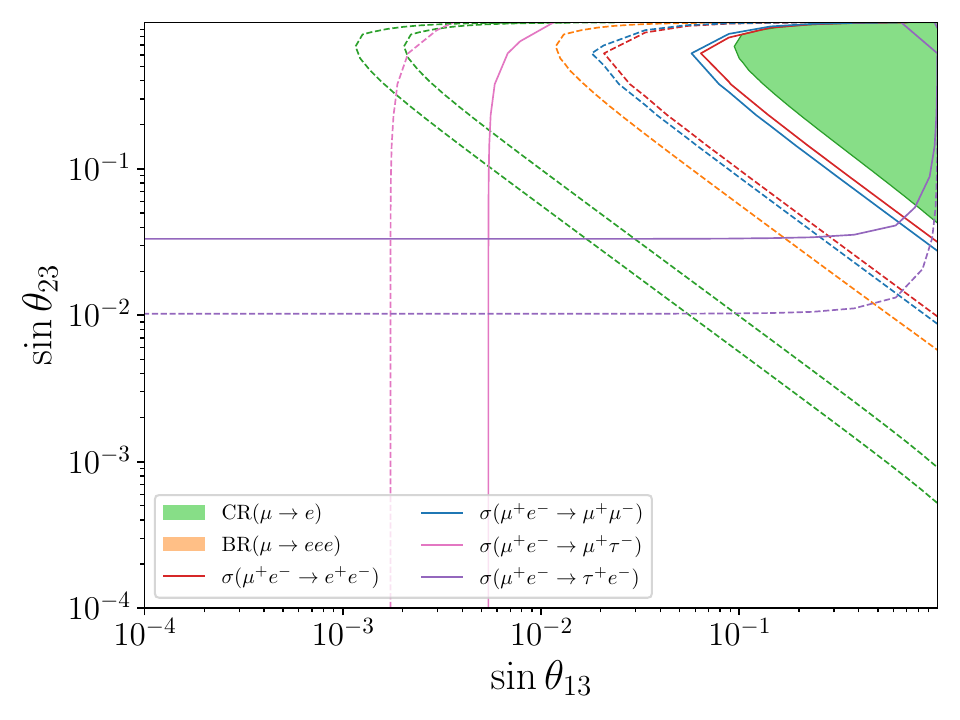}}
    \caption{Constraints from current experimental bounds (filled contours) and future sensitivities (dashed lines) from low-energy and $Z$-pole cLFV observables, together with the projected reach of a $\mu$TRISTAN collider from~\cite{2412.04331}, assuming luminosities of $100\:\mathrm{fb}^{-1}$ (solid lines) and $1\:\mathrm{ab}^{-1}$ (dashed lines). On each panel we vary two of the angles of $\mathcal V_2$ (see Eq.~(\ref{eqn:SVD}))  at a time with the other, as well as the off-diagonal $\eta_{ij}$, set to zero. The heavy spectrum has been fixed to $M_R = \mathrm{diag}(9,10,11)\:\mathrm{TeV}$.   
    The diagonal $\eta_{ii}$ have been set to their maximally allowed values as obtained in~\cite{Blennow:2023mqx}.}
    \label{fig:nd_si_vs_sj}
\end{figure}

\bigskip
To further demonstrate the impact of mixing amongst the heavy states, we now consider the  effect of simultaneously varying two mixing angles of $\mathcal V_2$ (the remaining one set to 0). The masses of the three heavy pseudo-Dirac pairs  are fixed to $M_R = \mathrm{diag}(9,\,10,\,11)\:\mathrm{TeV}$.
The results of this analysis are shown in Figure~\ref{fig:nd_si_vs_sj}, with the same colour code as above.
In the left plot we vary $\sin\theta_{12}$ and $\sin\theta_{13}$. 
As can be seen, there is a manifest interplay between the two mixings; having a large $\sin\theta_{13}$ leads to the suppression of the $\mu-e$ cLFV rates - induced by $\sin\theta_{12}$ - and vice versa.
Remarkably, the simultaneous presence of both further induces cLFV rates in the $\mu-\tau$ ``direction'': one can phenomenologically interpret this as a leakage of 
 flavour-violation from the $\mu-e$ and $\tau-e$ directions to the $\mu-\tau$ one.
This becomes all the more striking in the right plot of Figure~\ref{fig:nd_si_vs_sj}, where we vary $\sin\theta_{13}$ and $\sin\theta_{23}$.
Here, the simultaneous presence of both angles leads to sizeable rates for $\mu\to3e$ as well as $\mu-e$ conversion, albeit still in a sub-leading way with respect to the potential $\mu$TRISTAN sensitivities.

Finally, one can also consider the simultaneous presence of off-diagonal $\eta_{ij}$ and of non-vanishing mixing angles $\sin\theta_{ij}$ in $\mathcal V_2$.
This leads to more complicated mixing regimes in which it becomes challenging to disentangle the sources of flavour violation: between active and sterile states, or mixing among the heavy sterile states. The only exception to this is, as discussed, $\gamma$-penguin contributions. 

As previously argued, the $\gamma$ dipoles are to a very good approximation only proportional to one of the off-diagonal $\eta_{ij}$.
Consequently, contributions from $\gamma$-dipoles can be ``switched off'' by construction -- and thus it is possible to have sizeable rates for $\mu\to 3e$ and neutrinoless $\mu-e$ conversion in muonic atoms (induced by sizeable $\eta_{ii}$ and non-trivial mixing in the heavy sector) while having negligible rates for $\mu\to e\gamma$. 
The observation of such a pattern for the cLFV observables
could ultimately hint at a non-degenerate and non-trivially mixed heavy spectrum.

\section{Concluding remarks}
\label{sec:concs}
In this work we have considered an inverse seesaw mechanism with $3+3$ heavy sterile states.
Based on simple algebraic arguments, we developed new parametrisations of the Yukawa couplings that allow directly incorporating flavour misalignement stemming from  
the deviation from unitarity of the $3\times 3$ would-be PMNS block of the full unitary lepton mixing matrix. This deviation 
is conveniently 
encoded in the ``$\eta$-matrix'', which can be 
constrained by fits to low-energy data in a model
independent way.
Consequently, and by construction,
the new parametrisations we proposed here allow for a ``safe'' exploration of the ISS(3,3) viable parameter space, and one can systematically access different directions in ``flavour-violation space''.
We have then further studied this direct connection to low-energy data in a dedicated phenomenological analysis of numerous cLFV and lepton universality observables at low energies, at the $Z$-pole (at past and future lepton colliders), as well as at a possible $\mu^+ e^-$ $\mu$TRISTAN collider.

While $\mu\to e$ dedicated facilities will be able to set the strongest bounds on mixings associated with the first two generations, cLFV searches at $\mu$TRISTAN would improve the sensitivity to flavour violation in the  $e\tau$ and $\mu\tau$ sectors by several orders of magnitude with respect to dedicated searches at the $Z$-pole.
Our results reveal interesting synergies of the different cLFV observables and their constraining power to disentangle different sources of flavour violation: 
in particular, and as we have argued, they might allow distinguishing the flavour mixing between active and sterile neutrinos from mixing exclusively occurring amongst the heavy states.
This is in contrast to the Casas-Ibarra parametrisation in which the ``hidden'' parameters always have a severe impact on the Yukawa couplings and consequently on all observables at low energies.

We ultimately point out that in the absence of a signal in the $\mu\to e\gamma$ channel,
future measurements of sizeable rates for $\mu\to3e$ and neutrinoless $\mu-e$ conversion in muonic atoms could hint at the presence of a non-trivially mixed and non-degenerate heavy spectrum.

Although all parametrisations are formally equivalent (and do lead to the same physical results), here our goal was to develop a systematic access to phenomenological interesting regimes, exhibiting different flavour (and non-universality) features. 
Moreover, the analytical results of this work lead to a direct and simple connection to low-energy data, which allows for a ``safe'' implementation in Feynrules~\cite{Christensen:2008py,Alloul:2013bka}. 
This opens the door for phenomenological collider studies, which can be carried out while easily avoiding stringent constraints from low-energy cLFV and LFUV bounds.

\section*{Acknowledgements}
We are grateful to Emanuelle Pinsard for a careful reading of the manuscript.
JK is supported by the Slovenian Research Agency under the research grant No. N1-0253 and in part by J1-3013. \\
This project has received support from the European Union's Horizon 2020 research and innovation programme under the Marie Sk\l{}odowska-Curie grant agreement No.~860881 (HIDDe$\nu$ network) and from the IN2P3 (CNRS) Master Project, ``Hunting for Heavy Neutral Leptons'' (12-PH-0100).

\appendix
\section{Cholesky decomposition of $\eta$}\label{app:cholesky} 
Applying the Cholesky-Banachiwiecz algorithm on a $n\times n$ matrix of small order $n$ can be easily done by hand.
In the case of a complex hermitian $3\times 3$ matrix $A$ the Cholesky-decomposition $A = L L^\dagger$ is thus found as
\begin{eqnarray}
    L_{i,i} &=& \sqrt{A_{i,i} - \sum_{k=1}^{i-1} | L_{i,k}|^2}\,,\nonumber\\
    L_{i,j} &=& \frac{1}{L_{j,j}}\left(A_{i,j} - \sum_{k=1}^{j-1} L_{i,k}\,L_{j,k}^\ast\right)\quad \text{ for } i>j\,.
\end{eqnarray}
Furthermore, the inverse of the lower triangular matrix $L$ is easily found as
\begin{eqnarray}
    L^{-1} = \begin{pmatrix}
        \frac{1}{L_{1,1}} & 0 & 0\\
        -\frac{L_{2,1}}{L_{1,1}\,L_{2,2}} & \frac{1}{L_{2,2}} & 0\\
        -\frac{L_{2,2}\,L_{3,1}-L_{2,1}\,L_{3,2} }{L_{1,1}\,L_{2,2}\,L_{3,3}} & -\frac{L_{3,2}}{L_{2,2}\,L_{3,3}} & \frac{1}{L_{3,3}}
    \end{pmatrix}\,.
\end{eqnarray}


{
\small
\bibliographystyle{JHEP}
\bibliography{ISS-futurecLFV}
}

\end{document}